# CAUSAL PROPENSITY AS AN ANTECEDENT OF ENTREPRENEURIAL INTENTIONS IN TOURISM STUDENTS

Alicia Martín-Navarro, Félix Velicia-Martín,
José Aurelio Medina-Garrido, Ricardo Gouveia Rodrigues

This is the submitted version accepted for publication in the "International Entrepreneurship and Management Journal". The final published version can be found at: https://doi.org/10.1007/s11365-022-00826-1 We acknowledge that Springer holds the copyright of the final version of this work. Please, cite this paper in this way:

Martín-Navarro, A., Velicia-Martín, F., Medina-Garrido, J. A., & Rodrigues, R. G. (2023). Causal Propensity as an Antecedent of Entrepreneurial Intentions. International Entrepreneurship and Management Journal, 19(0123456789), 501–522.

## Abstract

The tourism sector is a sector with many opportunities for business development. Entrepreneurship in this sector promotes economic growth and job creation. Knowing how entrepreneurial intention develops facilitates its transformation into entrepreneurial behaviour. Entrepreneurial behaviour can adopt a causal logic, an effectual logic or a combination of both. Considering the causal logic, decision-making is done through prediction. In this way, entrepreneurs try to increase their market share by planning strategies and analysing possible deviations from their plans. Previous literature studies causal entrepreneurial behaviour, as well as variables such as creative innovation, proactive decisions and entrepreneurship training when the entrepreneur has already created his or her firm. However, there is an obvious gap at a stage prior to the start of entrepreneurial activity when the entrepreneurial intention is formed. This paper analyses how creativity, proactivity, entrepreneurship education and the propensity for causal behaviour influence entrepreneurial intentions. To achieve the research objective, we analysed a sample of 464 undergraduate tourism students from two universities in southern Spain. We used SmartPLS 3 software to apply a structural equation methodology to the measurement model composed of nine hypotheses. The results show, among other relationships, that causal propensity, entrepreneurship learning programmes and proactivity are antecedents of entrepreneurial intentions. These findings have implications for theory, as they fill a gap in the field of entrepreneurial intentions. Considering propensity towards causal behaviour before setting up the firm is unprecedented. Furthermore, the results of this study have practical implications for the design of public education policies and the promotion of business creation in the tourism sector. These policies should promote causal, proactive and creative behaviour in their entrepreneurship training. In this way, such policies would boost the entrepreneurial intention of individuals, which is an essential precursor to business creation.

## Introduction

Tourism is a field with many opportunities for business development. Many governments have recognised the importance of this sector and its contribution to the economy of the nations (Al-Jubari et al., 2019). Tourism is one of the largest industries in many countries, providing economic growth, foreign exchange earnings and jobs (Rauch & Hulsink, 2015). For this reason, governments promote and support entrepreneurship in this sector. Moreover, creating new businesses is an essential source of innovation, competitiveness and wealth for the rest of the business (Al-Jubari et al., 2019).

Entrepreneurial intention is a precursor to the creation of a firm (Ramos Rodríguea et al., 2019; Lechuga Sancho et al., 2020). The academic literature studies entrepreneurial intentions, often among university students (Anjum et al., 2018; Dewi & Sutisna, 2019; Hu et al., 2018; Laguía et al., 2019). Promoting entrepreneurial intention among tourism students can help transform their mental maps and orient them towards entrepreneurial behaviour. Therefore, exploring how to improve entrepreneurial intentions among tourism students is crucial for developing the industry now and in the future (Zhang et al., 2020).

When entrepreneurs start and manage their businesses they may develop different entrepreneurial behaviours, as suggested by effectual and causal logic (Sarasvathy, 2001). From a causal logic, decision-making is done through prediction. In this way, the entrepreneur tries to obtain the largest market share by planning strategies and controlling possible deviations from his plan. However, this paper focuses on the entrepreneurial intention of those individuals who have not yet created a company. For this reason, we will use the term causal propensity proposed by Martín-Navarro et al. (2021) to analyse this type of logic. The propensity for causal entrepreneurial behaviour, creativity, proactivity and learning programmes in entrepreneurship can influence the formation of entrepreneurial intention (Zampetakis et al., 2011; Kumar & Shukla, 2022; Jun et al., 2022; Leiva et al., 2021). In this sense, this paper aims to analyse the determinant power of these factors in the formation of entrepreneurial intention in the case of individuals who have not yet created a business.

As mentioned, the literature traditionally studies causal entrepreneurial behaviour in individuals who have already created their businesses. This paper fills a gap in the literature and adds value by studying the impact of individuals' causal propensity on their entrepreneurial intentions for the first time. In this context, these individuals have not yet created their businesses. However, the formation of their entrepreneurial intention is vital as it is a precursor of the entrepreneurial behaviour that eventually leads them to create their businesses.

To achieve the research objective, our paper is organised as follows. A literature review is conducted to develop our hypotheses in the following section. The methodology section presents a structural equation model application using SmartPLS software on a sample of 464 cases. Next, we explain the results obtained and the discussion. Finally, the conclusions, contributions, limitations and future lines of research are presented.

## Theoretical framework and hypothesis development

Enterprise creation plays a decisive role in the economy, especially in developing countries, increasing the level of employment and social growth (Farrukh et al., 2017). In this regard, many researchers have studied entrepreneurial intention as the stage before creating new businesses. The intention is a good indicator of future entrepreneurial behaviour (Anjum et al., 2018).

The literature has studied entrepreneurial intention and its determinants in depth. We find proactivity, creativity, program learning, and causal propensity among the multiple determinants analysed. The first two factors, proactivity and creativity, are two of the abilities that have the most significant influence on entrepreneurial intention (Hansen et al., 2011). On the one hand, proactive individuals seek to make changes or anticipate changes in the environment before they occur. Proactive people tend to take the initiative. These individuals bring about significant changes because they can influence the environment (Bateman & Crant, 1993).

On the other hand, creativity is the capacity to create, help entrepreneurs identify opportunities, and create new and innovative ideas (Schumpeter, 1934). Creative people are more likely to become entrepreneurs. Moreover, an entrepreneur can stimulate creativity among his or her employees by giving them more freedom and greater independence to innovate and generate original ideas (Kumar & Shukla, 2022).

In the entrepreneurial process, knowledge acquisition, as well as learning methods, have an impact on the identification and exploitation of new opportunities (Corbett, 2005). Entrepreneurship training can help students increase their alertness to opportunities by focusing on new trends and information (Baron, 2006). In this way, entrepreneurship training is vital as it links practical knowledge with the skills and confidence needed to succeed in business (Wilson et al., 2007). Setting up program learning in entrepreneurship can be costly. Therefore, in many cases, a link is established between companies and universities to support and create the right environment to grow the entrepreneurial spirit among university students (Nabi & Liñán, 2011).

Finally, the concept of causal propensity is based on the Effectuation Theory. This theory proposes that entrepreneurs can make decisions based on an effectual or causal logic. The effectual logic is based on creating opportunities from the individual's available resources. In contrast, the entrepreneur can also use causal logic. Causal logic is based on predicting the future and setting objectives to be achieved by that prediction. To achieve these objectives, entrepreneurs develop a process of planning, executing and controlling the planned activities. From this perspective, individuals use the information available to them and check that their strategy is being fulfilled. If this is not the case, the entrepreneurs study the possible causes of the deviations in their plan and take corrective measures (Sarasvathy, 2001). However, Effectuation Theory involves studying entrepreneurial behaviour when a company has already been created. It is not possible to analyse causal or effectual behaviours before when studying the determinants of entrepreneurial intention because these behaviours have not yet occurred and are not observable. In this sense, and following the work of Martín-Navarro et al. (2021), it would be possible to analyse individuals' propensity towards these behaviours. However, they have not yet developed these behaviours because they have not yet created their companies. For this reason, and in coherence with the objectives of this paper, we will use the term causal propensity. Causal propensity refers to the individual's tendency to make decisions and solve problems following causation logic.

As mentioned above, the academic literature has extensively studied entrepreneurial intentions and multiple determinants that influence entrepreneurial intentions. In our research, we propose four determinants as antecedents of entrepreneurial intentions. Among them program learning. The courses and training provided in educational institutions generate knowledge in students. The path commonly used to start a new project is developed from a causal logic. Thus, causal logic predominates as it is taught today in most business schools. According to the results obtained by Arvidsson et al. (2020), the vast majority of entrepreneurs, in their initial stage, tend to use causal logic as they need to set objectives within a framework of action that they have previously defined. As the entrepreneur gains experience, in a gradual process, he or she changes from a causal logic to an effectual one. Thus, from a causation approach, the organisation focuses on the most predictable aspects of an uncertain future and prediction is closely related to knowledge. The goal is clearly defined with a causal logic, and the available means are organised efficiently to reach that goal. In this way, the individual has a high level of technical knowledge that he or she has previously acquired in his or her formative stage (Héraud & Muller, 2016). Therefore, university education in entrepreneurship is linked to the causal propensity of the entrepreneur. All of the above makes it possible to state the following.

>**H1:** Program Learning is positively related to Causal Propensity.

The learning environment is essential to support creativity. Researchers have found that learning environments in which ideas are valued and mistakes are seen as a fundamental part of the learning process support creativity (Chan & Yuen, 2014). When teaching is in a group, its values influence the behaviours of its members, who influence each other to foster or constrain creativity (Peppler & Solomou, 2011). Individual learning does not foster creativity. However, the outcome is different in learning courses where students interact with the lecturer and with each other. Thus, with the lecturer's presence, a learning model is created in which university students considerably improve their creative ability. Thus, in a program learning model in which the lecturer acts as a facilitator of learning, the students become the main learning actors. Creativity becomes the main output of the learning process if a dynamic and interactive environment is used, together with the knowledge acquired by the students (Hardika et al., 2018). In this sense, Machali et al. (2021) found a strong correlation between entrepreneurship education and students' creativity. Valaei et al. (2017) found that learning was positively related to creativity within the context of companies within a sample of top managers. Thus, appropriate program learning facilitates the creativity of individuals.

Entrepreneurship training should not only focus on theoretical or methodological aspects, such as business plans. Increasingly, universities are striving to offer programmes to train future entrepreneurs in personal and managerial skills and in those personal traits that will help them to succeed. Because creativity is crucial in the entrepreneurial process, it is essential to cultivate it as a fundamental personality trait in individuals intending to become entrepreneurs (Hu et al., 2018). Concerning the proactive behaviours of individuals, some researchers have argued that proactivity is influenced by behaviours learned through training initiatives (Fay & Frese, 2001; Kabanoff & Bottger, 1991). Thus, proactivity as a set of action-directed behaviours is influenced by personality traits such as creativity (Van Veldhoven & Dorenbosch, 2008). The relationship between the creativity of individuals and their proactive capacity allows us to indicate. The above arguments establish the following relationships.

>**H2:** Program Learning is positively related to creativity.

>**H3:** Creativity is positively related to proactivity.

As discussed above, proactive entrepreneurs can exhibit two distinct, non-exclusive behaviours. One behaviour is related to deliberately generating environmental changes to generate opportunities that did not previously exist. Nevertheless, also proactive are those entrepreneurs who foresee the changes that the environment has in store for them and are prepared even before this change takes place. These proactive individuals detect good opportunities much earlier than others, develop better strategies (Seibert et al., 2001), and develop better strategic vigilance (Yeşilkaya & Ylldlz, 2022).

Strategic monitoring is a tool for predicting the future that involves the ability to acutely evaluate information regarding competitors and customers (Yeşilkaya, 2015). Thanks to this ability, entrepreneurs can anticipate and plan what they will offer to their customers (Mahmoud & Mahdi, 2019), anticipate the steps to take in the face of competition, convert threats in the environment into opportunities and establish strategies to adapt to change (Alshaer, 2020; Yeşilkaya, 2015). Furthermore, establish strategies to adapt to change (Yeşilkaya & Ylldlz, 2022).

Therefore, strategic vigilance and the ability to detect opportunities enable proactive individuals to rationally plan to take advantage of these opportunities and set goals to be achieved within a causal logic. Similarly, proactivity will support individuals' causal propensity because the more significant the ability to spot opportunities before others, the easier it is to plan how to take advantage of them. Therefore, as indicated below, individuals' proactivity can be connected to a potential causal behaviour, i.e. causal propensity.

>**H4:** Proactivity is positively related to Causal Propensity.

The uncertainty that entrepreneurs face when creating new ventures involves factors that cannot be predicted and unknown variables. Causal logic understands that the future can be predicted if sufficient information is obtained, for example, through market research. In the causal process, a lot of time and resources are invested in developing a plan to succeed in an existing market (Sarasvathy, 2008; 2001). When individuals set up a new enterprise, they pursue performance and plan how to reach that goal. From a causal approach, both issues are fundamental to forming entrepreneurial intentions (Dutta et al., 2015). The causation approach assists the entrepreneur in developing a new business (Sarasvathy, 2001b) and gathering resources efficiently, working under the entrepreneur's strategy (Delmar & Shane, 2004). In this regard, Li et al. (2020)found that causal logic positively affects entrepreneurial behaviour in a sample of managers in Pakistan. By analogy, if the individual is not yet entrepreneurial, causal propensity should positively impact entrepreneurial intentions, as hypothesised below.

**H5:** Causal Propensity is positively related to Entrepreneurial Intention.

Proactive individuals can propose constructive changes that can lead to more appropriate systems. On the other hand, passive and inactive individuals are left to the circumstances as they occur (Kumar & Shukla, 2022). A proactive personality is a key driver for achieving an advantageous outcome in challenging situations (Gupta & Bhawe, 2007). Proactivity is a crucial element of foresight. It is linked to the management and entrepreneurial behaviour. It allows exploring uncertainty while simultaneously executing actions that can influence the future (Djuricic & Bootz, 2019). It is precisely proactivity that helps the entrepreneur to survive in a turbulent environment full of uncertainties (Godet, 1993). Sidratulmunthah et al. (2018) found empirical evidence of the relationship between proactivity and entrepreneurial intentions in a sample of 306 female students from business universities in Pakistan.

Similarly, Huston (2018) found that proactivity and innovativeness were antecedents of entrepreneurial intention in a sample of Doctor of Pharmacy students in a pharmacy practise management course at the University of Georgia. Castillo & Fischer (2019) also found that proactivity influenced the desire to be entrepreneurial in a sample of individuals with disabilities. In summary, proactive individuals will be more likely to have entrepreneurial intentions than others, as expressed below.

**H6:** Proactivity is positively related to Entrepreneurial Intention.

A growing body of research links creativity with the generation of new and valuable ideas (Amabile, 1996) with entrepreneurship (Mahmood et al., 2018). Creativity enables entrepreneurs to innovate, generate ideas and identify opportunities. In this way, creativity is consolidated as a critical component of entrepreneurship (Schumpeter, 1934). Thus, the more creative individuals are, the more likely they are to be entrepreneurs (Hamidi et al., 2008). Mahmood et al. (2018) showed that the creativity of MBA graduates with experience in the labour market was an antecedent of entrepreneurial intention.

Similarly, Zampetakis et al. (2011) also found a positive relationship between creativity and entrepreneurial intentions in a sample of 180 undergraduate students in England. This relationship was also supported in a sample of 484 undergraduate students in the work of Kumar & Shukla (2022). With these results, we can posit that creativity is positively related to entrepreneurial intentions.

**H7:** Creativity is positively related to Entrepreneurial Intention.

Program learning refers to the learner's knowledge during training courses (Souitaris et al., 2007). Numerous studies have found a positive relationship between program learning and entrepreneurial intentions. Rauch & Hulsink (2015) One study found that the entrepreneurship training program for undergraduate students at a university in the Netherlands positively influenced their entrepreneurial intentions. Similarly, empirical evidence that entrepreneurship education positively affects entrepreneurial intentions was found in a sample of university students from several Latin American countries (Leiva et al., 2021). The relationship between entrepreneurship education and entrepreneurial intentions was also positive in a sample of fourth-year undergraduate business administration students at the University of Granada in Spain (González-López et al., 2019). The above sources of evidence allow the following research hypothesis to be put forward.

**H8:** Program Learning is positively related to Entrepreneurial Intention.

In environments of uncertainty, entrepreneurs adopt more causal logic to take advantage of opportunities that arise the more information and knowledge they acquire about the evolving environment (Yao et al., 2013). The causal orientation involves the creative exploitation of opportunities and the development of business plans (Chandler et al., 2011; Sarasvathy & Dew, 2008). In addition, creativity is a helpful resource for recognising when it is necessary to make established plans more flexible (de Vasconcellos et al., 2019), planning being the basis of causal logic. Creativity also impacts the propensity towards causal logic in decision-making at an early stage where the individual is not yet entrepreneurial. The creativity that makes it possible for entrepreneurs to seize business opportunities fuels their potential ability to generate better business plans to take advantage of them. Based on the above arguments, the following hypothesis captures the relationship between creativity and causal propensity.

**H9:** Creativity is positively related to Causal Propensity.

Based on the above arguments and the connections discovered above, we establish the research model shown in Figure 1.

**Figure 1. Research model**

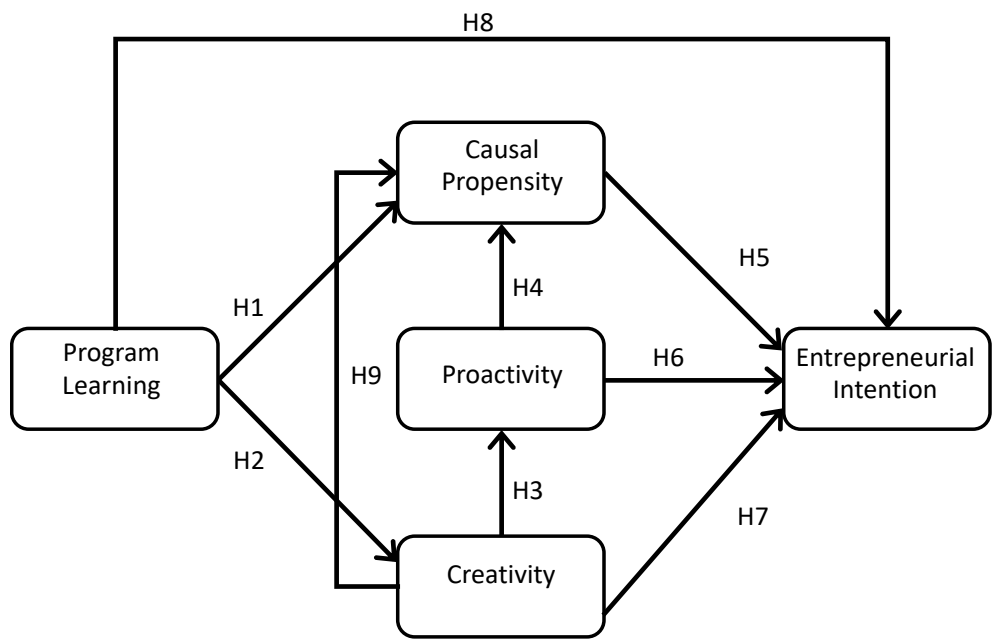

# Methodology

## Data collection

The sample size was n=464 respondents. The sample was obtained with the voluntary collaboration of the students of the degree course in Tourism of the universities of southern Spain. They were interviewed employing a self-administered questionnaire using non-probabilistic sampling. The questionnaire was distributed between April and July 2021. The questionnaire items were measured with a seven-point Likert scale: one strongly disagreed, and seven strongly agreed.

The main socio-demographic characteristics of the sample are: in relation to gender, 71.02% of the respondents are female compared to 28.98% male. Concerning age, 56.04% of the respondents are between 18 and 21 years old, 37.58% are between 22 and 25 years old, and the rest are over 25 years old. The 97.4% are Spanish nationals, and only 2.6% are foreigners. Regarding entrepreneurial tradition, 31.82% of those surveyed belong to a family whose parents have had or have had their own business. Regarding their employment situation, 34.20% have worked or are currently working as employees, compared to 46.84% who have never worked; 18.96% have worked or are currently working as self-employed.

## Measurement scales

The constructs of our model explain indicators that are highly correlated and interchangeable. It is, therefore, a composite model with reflective indicators.

Causal propensity is an individual's tendency towards predictive logic based on his or her knowledge in making decisions before undertaking a business venture (Martín-Navarro et al., 2021). We used seven indicators from Martín-Navarro et al. (2021) to measure it. According to Amabile (1996), creativity is the ability to create new and valuable ideas. Creativity was adapted from Zampetakis (2008) and consisted of four items. Program Learning provides the context for learners to acquire skills and behaviours to create value in enterprises (Gundry et al., 2014). This construct was adapted from GUESSS questionnaire (Lechuga Sancho et al., 2020) and consists of five items. The individual's certainty to plan and create a new enterprise in the future (Thompson, 2009) defines Entrepreneurial Intentions. It comprises six items adapted from Liñán & Chen (2009). Moreover, proactivity, as the individual's active efforts to effect changes in his or her environment (Zampetakis, 2008), is measured with ten items from Seibert et al. (2001). All indicators are listed in Appendix 1.

## Data analytics

In order to test the hypotheses, we used a structural equation model (Hair et al., 2014). The first objective of PLS-SEM is to determine the amount of variance explained in the endogenous variables of a structural model and explain the relationships posited in the model (Hwang et al., 2020). We use the Smart PLS 3.0 software to carry out the data analysis, following the indications in Ringle et al. (2015). The evaluation of the research model was carried out in two distinct stages. In order to determine the validity and reliability of the constructs, firstly, we carried out an analysis of the measurement model. Secondly, in order to conclude the relationships, we carried out an analysis of the structural model.

# Results

## Analysis of the measurement model

Common method bias (CMB), in the context of research using PLS-SEM, is a phenomenon that could be caused by the measurement method, typically Likert-type questionnaires (Kock, 2015).

CMB is a serious threat because bias can affect the findings due to systematic errors (Schwarz et al., 2017). In this research, we have tried to avoid CMB in the research development and applied the statistical procedure suggested by Kock (2015) to detect it. When VIF coefficients are higher than 3.3, it indicates collinearity so that CMB may contaminate the model. Table 1 shows the VIF coefficients that are clearly below the 3.3 limit.

**Table 1. Full collinearity VIFs**

| Variables | Program Learning | Causal Propensity | Creativity | Proactivity |
|---|---|---|---|---|
| **VIF** | 1.37 | 1.30 | 1.62 | 1.80 |

The PLS-SEM method was used for the analysis. PLS is a recommended method for studying latent construct models made up of composite (Rigdon, 2016). Individual reliability, construct reliability, discriminant validity and convergent validity were assessed. Reliability ensures that the measurement produces consistent results, and validity ensures that the indicators of a construct measure the construct they are intended to measure and not another one (Hair et al., 2011).

The individual reliability of each item was analysed, and items with loadings below 0.707 were eliminated (Carmines & Zeller, 1979). Subsequently, construct reliability is analysed through Dijkstra-Henseler rho_A in all cases with values above 0.7 (Dijkstra & Henseler, 2015). Composite reliability should be above 0.8 (Nunnally & Bernstein, 1995). Convergent validity is tested through the Average Variance Extracted (AVE). The values must be greater than 0.5 (Fornell & Larcker, 1981). Table 2 shows that the established requirements are satisfied.

**Table 2. Reliability and convergent validity**

| Construct/Indicator | Loads | rho_A | Compound reliability | AVE |
|---|---|---|---|---|
| **Program Learning (PL)** | | 0.913 | 0.935 | 0.742 |
| Learn01 | 0.859 | | | |
| Learn02 | 0.875 | | | |
| Learn03 | 0.872 | | | |
| Learn04 | 0.845 | | | |
| Learn05 | 0.856 | | | |
| **Causal Propensity (CP)** | | 0.739 | 0.846 | 0.648 |
| Caus05 | 0.850 | | | |
| Caus06 | 0.816 | | | |
| Caus07 | 0.745 | | | |
| **Creativity (CREA)** | | 0.838 | 0.891 | 0.672 |
| Crea01 | 0.834 | | | |
| Crea02 | 0.746 | | | |
| Crea03 | 0.863 | | | |
| Crea04 | 0.831 | | | |
| **Proactivity (PROA)** | | 0.852 | 0.889 | 0.615 |

| | | | |
|---|---|---|---|
| Proact05 | 0.706 | | |
| Proact07 | 0.813 | | |
| Proact08 | 0.795 | | |
| Proact09 | 0.807 | | |
| Proact10 | 0.809 | | |
| **Entrepreneurial Intention** | | 0.898  0.934 | 0.825 |
| EI01 | 0.904 | | |
| EI03 | 0.935 | | |
| EI05 | 0.884 | | |

Discriminant validity determines whether a particular construct is different from others. In our study, discriminant validity was found (Table 3). For this purpose, the criteria of Fornell-Larcker and the Heterotrait-Momonotrait Ratio (HTMT) (Henseler et al., 2016) were applied. There is discriminant validity given that the HTMT values are below 0.9 (Gold et al., 2001). The results confirm that the measurement model is valid and reliable.

**Table 3. Discriminant validity**

| | Fornell-Larcker | | | | | HTMT | | | |
|---|---|---|---|---|---|---|---|---|---|
| | PL | CP | CREA | PROA | EI | PL | CP | CREA | PROA |
| **PL** | **0.861** | | | | | | | | |
| **CP** | 0.344 | **0.805** | | | | 0.423 | | | |
| **CREA** | 0.408 | 0.365 | **0.82** | | | 0.464 | 0.466 | | |
| **PROA** | 0.405 | 0.422 | 0.581 | **0.784** | | 0.454 | 0.528 | 0.686 | |
| **EI** | 0.379 | 0.292 | 0.315 | 0.435 | **0.908** | 0.418 | 0.358 | 0.362 | 0.495 |

## *Structural model*

After verifying reliability and validity, the structural model is evaluated to test the hypotheses and relationships between the constructs proposed in the research model (Saura et al., 2020). To evaluate the structural model (estimation of path loadings and $R^2$), bootstrapping with 5,000 resamples is used to test the proposed hypotheses (Hair et al., 2011).

Standardised path coefficients (β) explain the size of the predictor variables' contribution to the endogenous variables' variance (Palos-Sanchez et al., 2021). The results of the path loadings coefficient are shown in Table 4. These results show that eight of the nine relationships proposed in the theoretical model are significant. Hypothesis H7 was not supported. That is, the relationship between creativity and entrepreneurial intentions is not significant.

**Table 4. Results of significance tests of the coefficients of the structural model**

| Hypothesis | β (Standard Path Coeff.) | T-Statistics | P-Values | CI | Sig | |
|---|---|---|---|---|---|---|
| H1: PL→CP | 0.18 | 3.466 | 0.000 | (0.094;0.264) | Yes | *** |
| H2: PL→CREA | 0.408 | 9.285 | 0.000 | (0.336;0.481) | Yes | *** |
| H3: CREA→PROA | 0.581 | 16.411 | 0.000 | (0.522;0.636) | Yes | ** |
| H4: PROA→CP | 0.271 | 4.882 | 0.000 | (0.178;0.362) | Yes | *** |

| H5: CP→EI | 0.082 | 1.676 | 0.047 | (0.002;0.161) | Yes | * |
| H6: PROA→EI | 0.3 | 5.21 | 0.000 | (0.206;0.394) | Yes | *** |
| H7: CREA→EI | 0.02 | 0.377 | 0.353 | (-0.071;0.104) | No | n.s. |
| H8: PL→EI | 0.221 | 4.233 | 0.000 | (0.135;0.308) | Yes | *** |
| H9: CREA→CP | 0.134 | 2.419 | 0.008 | (0.043;0.226) | Yes | ** |

Note: Significant at p*<0.05; p**=<0.01 and p***=<0.001

The model's explanatory power can be measured by studying the $R^2$ value of the dependent variable EI. The $R^2$ indicates the variance explained by the exogenous variables (Ramírez-Correa et al., 2019). Figure 2 shows the explanatory power of the other constructs. $R^2$ values above 0.67 are considered high, between 0.67, and 0.33 moderate, between 0.33 and 0.19 weak and values below 0.19 are unacceptable (Hair et al., 2014). In this case, Entrepreneurial Intention and Causal Propensity present a weak $R^2$, while the values for proactivity are moderate.

**Figure 2. Quality of the measurement model and the structural model**

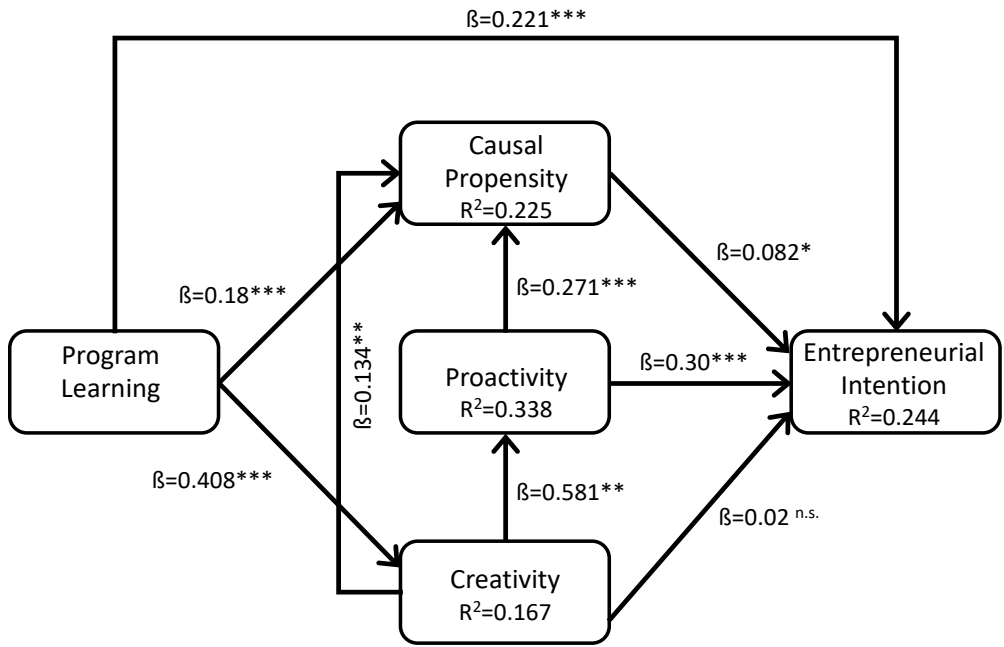

The global model fit is performed using the values of the standardised root mean square residual (SRMR). A value of 0 would represent a perfect fit, although values below 0.08 present a good fit (Henseler et al., 2014). In our model, SRMR presents a value of 0.055 lower than the reference value, so our model presents a good fit.

## *Predictive model*

Assessing the predictive power of a model is fundamental in social science research (Hair, 2021). In this respect, PLS-SEM facilitates two critical types of analysis. First, it explains the research model. Something we have already developed in the previous stages. Moreover, secondly, it performs the prediction of the model (Joreskog & Wold, 1982).

In the past, researchers have interpreted the coefficient of determination ($R^2$) to predict the values of individual cases in a sample. Nevertheless, the R-value$^2$ only assesses the explanatory power of a model and not the out-of-sample predictive power concerning new cases (Shmueli, 2010). Assessing the predictive power of a model in a given sample involves measuring its performance for data other than those in the original sample (Cepeda Carrión et al., 2016). In this sense, we analyse the predictive power of our model with $Q^2$ (Geisser, 1974; Stone, 1974). $Q^2$ is a metric provided by PLSpredict that is frequently used in this context and is obtained through the blindfolding procedure (Chin, 1998).

The model's predictive power is analysed with the PLSpredict technique, which tests the model's generalizability to other populations. We obtained results from the variables RMSE (R Mean Square Error), MAE (Mean Absolute Error), and Q_predict$^2$. The first two are composite score-based prediction errors. Moreover, concerning the third one, for the model to have predictive power, the value of $Q^2$ must be greater than zero (Shmueli et al., 2019). In our case, all the values in the last column are positive (see Table 5). Similarly, we obtained positive values for RMSE and MAE, showing good predictability (Woodside, 2013). Therefore, this research demonstrated the proposed model's predictive validity, or out-of-sample predictability, to predict the values of new cases.

**Table 5. Evaluation of PLS predictions**

|  | RMSE | MAE | $Q^2$ |
|---|---|---|---|
| Causal Propensity | 0.949 | 0.754 | 0.11 |
| Creativity | 0.922 | 0.719 | 0.161 |
| Proactivity | 0.936 | 0.741 | 0.133 |
| Entrepreneurial intention | 0.932 | 0.764 | 0.137 |

# Discussion

In our paper, we have analysed four antecedents of the entrepreneurial intention of Tourism students at the University of Cadiz and the University of Seville (Spain): program learning, creativity, proactivity and causal propensity. We have also analysed different relationships between these variables. These relationships allow us to configure a theoretical model from which nine research hypotheses are derived. The results of the empirical study have confirmed eight of the nine hypotheses put forward in the research model.

Thus, we have found empirical evidence for the effect of program learning on causal propensity, so H1 is supported. Our results are consistent with those of Arvidsson et al. (2020). In this sense, individuals who have never started a business may develop causal propensity due to the entrepreneurship training they have received. The results of the study also support H2. Therefore, we can state that program learning positively affects student creativity. This result is in line with the findings of Machali et al. (2021) y Valaei et al. (2017), who also demonstrated this relationship in their research.

On the other hand, there are few empirical studies on the effect of creativity on proactivity in the academic literature (Zampetakis, 2008). In our work, we see that the results confirm that creativity affects proactive behaviour (H3 is supported). Similarly, although they did not directly study the effect of creativity on proactivity, Bourmistrov and Åmo (2022) found that creativity was related to some components of individuals' foresight. Similar to our results, Zampetakis (2008) found that creativity is an important variable that affects individuals' proactivity towards their entrepreneurial intentions.

About H4, in most of the works reviewed, the literature considers that proactive behaviour is typical of an effectual decision-making logic (Chen et al., 2021). Considering this gap, our research proposed that proactivity could also be a determinant of causal logic and, more specifically, of the propensity towards causal behaviour. The results obtained confirm the relationship between individuals' proactivity and causal propensity. Therefore, H4 is supported. These results are consistent with Yeşilkaya and Yildiz (2022), for whom proactive individuals detect good opportunities much earlier than others and develop better strategic vigilance. This strategic vigilance gives proactive individuals the ability to rationally plan how to seize these opportunities (Yeşilkaya,2015; Alshaer,2020) and set goals to achieve within a causal logic, anticipating and planning what they will offer to their customers (Mahmoud & Mahdi, 2019) and how they will adapt to changes in the environment (Yeşilkaya & Yildiz, 2022).

Hypotheses H5, H6 AND H7 of the model posited in this study propose that causal propensity, proactivity and creativity are antecedents of entrepreneurial intentions. The first two of these relationships are supported (H5 and H6). However, the third, H7, is not satisfied as the relationship between creativity and intentions is not significant. Our results are in line with those of Li et al. (2020), who also found that causal behaviour affected intentions to start a business. Similarly, our findings support the results of Sidratulmuntah et al. (2018), Huston (2018), and Castillo & Fischer (2019), who found that proactivity affects intentions. However, in our study, creativity is not a determinant of entrepreneurial intentions. This result is contrary to those of Kumar & Shukla (2022) y Zampetakis et al. (2011). The previous authors did find empirical evidence for the relationship between creativity and entrepreneurial intentions. Similarly, Zampetakis (2008) demonstrated that proactivity and creativity affect intentions among the Greek students who participated in their research, mediated by the desire to be entrepreneurial.

Many researchers have found the positive effect of program learning on entrepreneurial intentions (González-López et al., 2019; Leiva et al., 2021; Rauch & Hulsink, 2015). Our results confirm these findings as H8 is supported. Therefore, our sample finds that program learning is an antecedent of entrepreneurial intentions. Finally, we can affirm that creativity determines causal propensity since H9 is also satisfied. These results are consistent with the arguments of Chandler et al. (2011) and Sarasvathy and Dew (2008). These researchers claimed that creative ideas were part of the causal logic. By analogy, our results show that they are also antecedents and part of causal propensity.

## Conclusions

There are many studies on students' entrepreneurial intentions (Barba-Sánchez et al., 2022; Leiva et al., 2021; Wang et al., 2022). Nevertheless, this type of study in the tourism sector is especially relevant due to the opportunities this industry offers for business development and generating wealth and employment. Our study analysed a sample of undergraduate tourism students at universities in southern Spain. This work proposed a research model with four variables as antecedents of entrepreneurial intentions (program learning, creativity, proactivity and causal propensity). The inclusion of causal propensity in this model is innovative as it is the first time it has been studied as an antecedent of entrepreneurial intentions. This variable refers to the tendency towards the causal logic of people who have not yet started their business.

The proposed theoretical model established nine research hypotheses. The results support eight of these hypotheses, and only one was rejected. Thus, three of the four determinants proposed in the research influence entrepreneurial intentions. These determinants are program learning, proactivity and causal propensity. However, according to the results of this study, there is no evidence that creativity directly influences entrepreneurial intention.

The predictive capacity of the theoretical model established in this work is also remarkable. We have evaluated our model with PLSpredict, and the result shows that it has a high predictive ability, which indicates that similar results will be obtained in other samples with new cases.

## *Contributions*

Given the results obtained, our work has important implications for academic research and practice. First, this study adds value to the literature on entrepreneurial intention as it is the first time that the impact of the causal propensity of individuals who have not yet started their business on their entrepreneurial intentions has been tested. Secondly, the sequential mediating effect of creativity, proactivity and causal propensity on the relationship between program learning and entrepreneurial intention has been tested.

Our results also have implications for the educational community, policymakers, and other agents working to promote self-employment. The results of this study show that to promote entrepreneurial intentions, it is necessary to offer training in entrepreneurship that develops students' creativity and proactivity. In addition, students need to learn how to make business plans in which they plan the opportunities to be seized and the objectives to be achieved. By analogy, employers who wish to support intrapreneurship among their employees should also consider the importance of training that fosters creativity, proactivity and causal propensity to generate entrepreneurial intentions that encourage the development of projects within the organisation or in spin-offs.

## *Limitations and future agenda*

This work has some limitations that also present opportunities for future research. First, the sample used in this study comprises tourism students from two universities that are very close geographically, so their responses do not differ too much. In this sense, we propose that the researchers carry out this study with students from other university degrees and in different cultural contexts and countries. Second, this paper only studies entrepreneurial intentions prior to the creation of the company. The theoretical model proposed in this paper could be completed by including the next stage of the entrepreneurial process, that is, the launch of the start-up company. Third, the proposed theoretical model could be extended with other personal factors such as effectual propensity, passion, optimism, or risk aversion, among other variables.

Moreover, fourth another limitation is that the causal propensity construct lost several items in the factor analysis that were present in the original measurement scale proposed by Martín-Navarro et al. (2021). For that, additional effort is required to confirm the construct's consistency. It would be interesting to test the consistency of the construct in other samples in future research.